\documentclass[graybox]{svmult_gam}

\usepackage{mathptmx}       % selects Times Roman as basic font
\usepackage{helvet}         % selects Helvetica as sans-serif font
\usepackage{courier}        % selects Courier as typewriter font
\usepackage{type1cm}        % activate if the above 3 fonts are
\usepackage{makeidx}         % allows index generation
\usepackage{graphicx}        % standard LaTeX graphics tool
\usepackage{multicol}        % used for the two-column index
\usepackage[bottom]{footmisc}% places footnotes at page bottom
\usepackage{natbib}
\usepackage{journals}
\setcitestyle{numbers,square}
\makeindex             % used for the subject index
\begin{document}

\title*{Deprojecting the quenching of star formation in and near clusters}
\author{G. A. Mamon, S. Mahajan and S. Raychaudhury}
\institute{G. A. Mamon \at IAP (UMR 7095: CNRS \& UPMC), Paris, France, \email{gam@iap.fr}
\and S. Mahajan \& S. Raychaudhury \at School of Physics \& Astronomy, Univ. of Birmingham, UK}

%\dedication{In memory of Chris Moss, who passed away 12 May 2010}

\maketitle

% Too much empty space in the original style file!
\vskip-1.2truein

\abstract{Using H$_\delta$ and D$_n$4000 as tracers of 
recent or ongoing efficient star formation, we analyze the fraction
of SDSS galaxies with recent or ongoing efficient star formation (GORES) in the
vicinity of
268 clusters.
We confirm the well-known segregation of star formation, and using Abel
deprojection, we find that the fraction of
GORES increases linearly with physical radius and then
saturates. 
Moreover,
we find that the
fraction of GORES is modulated by the absolute line-of-sight velocity
(ALOSV): at all projected radii, higher fractions of GORES are found in
higher ALOSV galaxies.
We model this velocity modulation of GORES fraction using the particles in a
hydrodynamical cosmological simulation, which we classify into virialized,
infalling and backsplash according to their position in radial phase space at $z=0$.
Our simplest model, where the GORES fraction is only a function of class does
not produce an adequate fit to our observed GORES fraction in projected phase
space.
On the other hand, assuming that in each class the fraction of GORES rises
linearly and then saturates, we are able to find well-fitting 3D models of
the fractions of GORES. In our best-fitting models, in
comparison with 18\% in the virial cone and 13\% in the virial sphere,
GORES respectively
account for
34\% and 19\% of the infalling and backsplash galaxies, and as much as 11\%
of the virialized galaxies, possibly as a result of tidally induced star
formation from galaxy-galaxy interactions.
At the virial radius, the fraction of GORES of the backsplash population is
much closer to that of the virialized population than to that of the
infalling galaxies. This suggests that the quenching of efficient star
formation is nearly complete in a single passage through the cluster.}

\section{Introduction}

It is well known that the cluster environment affects the physical properties
of galaxies, as there is a segregation with projected radius of morphology
\citep{Dressler80}, color (e.g. [\citealp{Balogh+04_ApJL}]), luminosity
(e.g. [\citealp{ABM98}]) and spectral indices such as the equivalent width of
      [OII] or H$_\alpha$ (e.g. [\citealp{Balogh+04_MN,Haines+06}]).
At the same time, there are indications of velocity segregation of luminosity
\citep{Biviano+92}
and of star formation efficiency: emission-line
galaxies tend to span a wider dispersion of velocities than galaxies without
such lines \citep{Biviano+97}. 
In fact, it was long known that spiral galaxies in clusters span a wider distribution of
velocities than ellipticals and S0s (e.g. [\citealp{MD77}]). However, 
those spirals without emission lines and that are not morphologically
disturbed span the same velocity distribution as the early type galaxies \citep{Moss06}.
Relative to passive galaxies (without
H$_\alpha$ in emission), the trend for higher velocity dispersions of galaxies with
emission (H$_\alpha$) lines found by \citep{Biviano+97} is reversed
outside the virial radius \citep{RGKD05}.

In the present work \citep{MMR11}, we take advantage of the large statistics
of the Sloan Digital Sky Survey (SDSS) to better quantify the velocity
modulation of the radial segregation of the diagnostics of star formation
efficiency.
We then model the SDSS observed fractions of Galaxies with Ongoing or Recent
Efficient Star Formation (GORES), with the help of a cosmological simulation.

\section{Observed Velocity Modulation}

We select clusters of at least 15 galaxies from the SDSS-DR4 group catalog of
\citep{Yang+07}, 
with criteria $0.02\leq z\leq0.12$, $M_{180} > 10^{14} M_\odot$ and with at
least 12 $M_r<-20.5$ galaxies within $R_{180}$.
Around these 268 clusters, we select galaxies in a slightly wider redshift
range, lying within $2\,R_{100}=2.6\,R_{180}$ from the cluster in projected
space,
with absolute line-of-sight velocity relative to the cluster mean (ALOSV)
within $3\sigma_\upsilon$,
 with absolute magnitude $M_r < -20.5$, and with angular effective radius
 $\theta_{\rm eff} \leq 5''$ to avoid large galaxies for which the SDSS
 fibers only span the inner region.
This selection yields 19$\,$904 galaxies and we also select 21$\,$000 field galaxies
that lie at least 10 Mpc in projection away from any cluster (corresponding
to typically $6.7\,R_{\rm v}$, where the virial radius $R_{\rm v}\equiv R_{100}$).

\begin{figure}
\centering
\includegraphics[width=0.49\hsize]{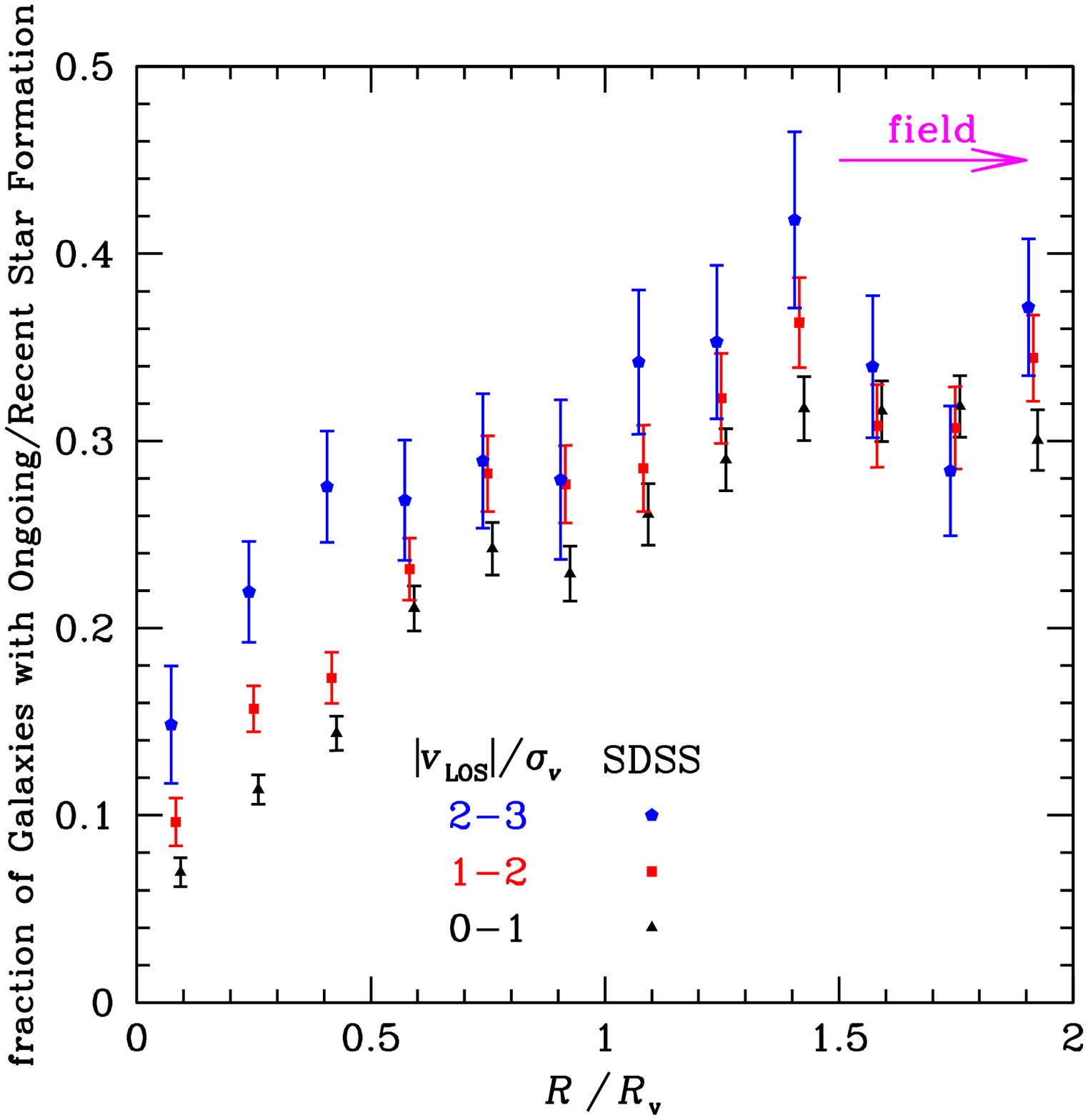}
\includegraphics[width=0.49\hsize]{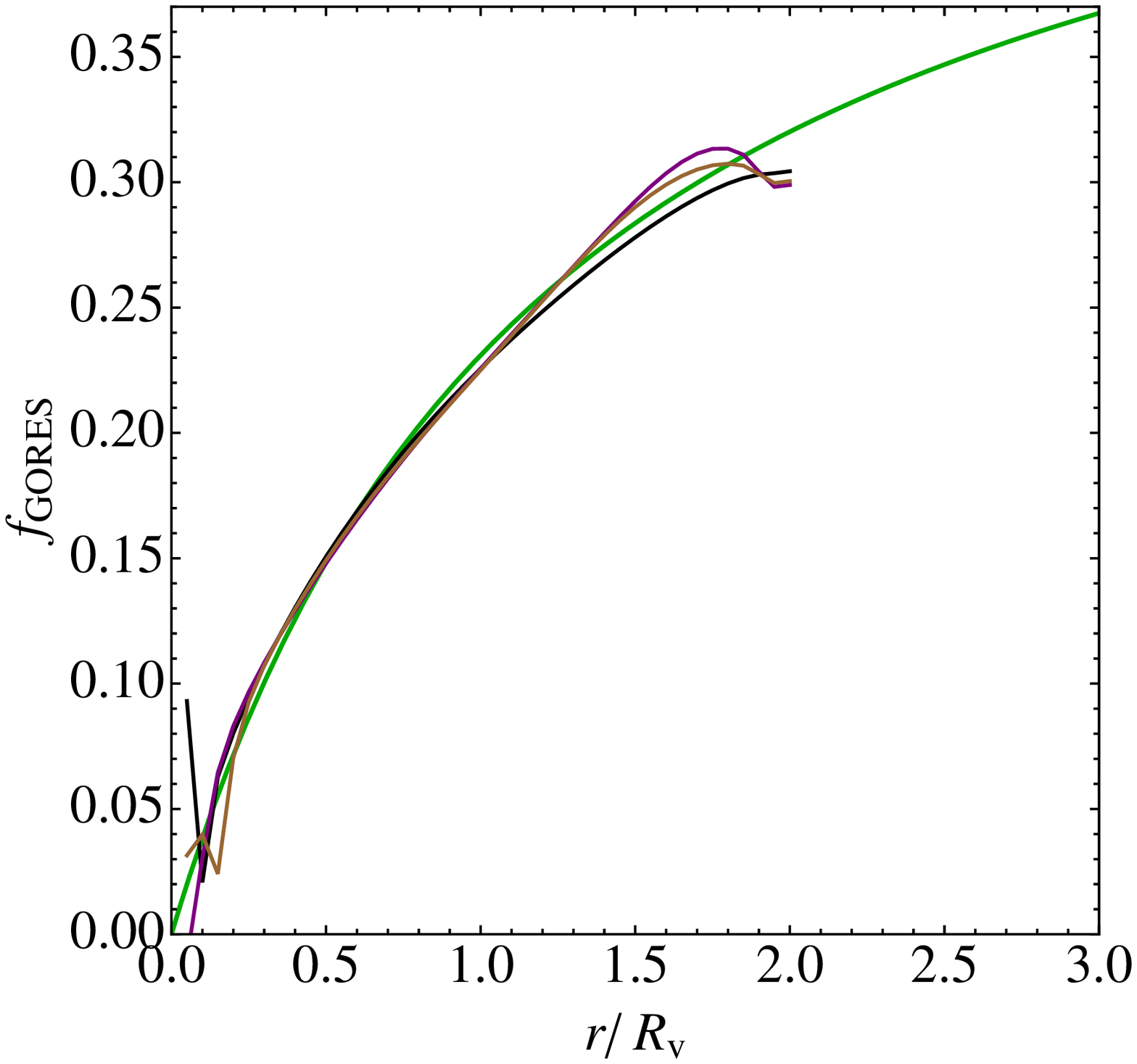}
\caption{\emph{Left}: Observed fraction of GORES in three bins of ALOSV.
\emph{Right}:  Deprojected fractions of GORES.
 \emph{Thin curves} show polynomial fits of the log surface density vs.
%%  $R$ (orders
%%   4 and 6 in \emph{black} and \emph{red}, respectively) and
  $\log R$ (orders 3, 4, and 5). 
The \emph{longer thick curve} is %%  the fit of
Eq.~(\ref{modelfit}) with
$f_0=0.52$ and
$a = 1.26\,R_{\rm v}$, obtained by a $\chi^2$ fit to the order 4 polynomial
fit of $\log \Sigma$ vs. $\log R$ with $r$ linearly
spaced between 0.05 and $2\,R_{\rm v}$.
\label{gGORES}}
\end{figure}

We identify GORES as galaxies with both H$_\delta> 2\,\rm \AA$ and D$_n$4000
$<$ 1.5 (age less than 1 Gyr for most galaxies, [\citealp{Kauffmann+03}]). 
The symbols in the left panel of Fig.~\ref{gGORES} illustrate how the observed
fraction of GORES increases with projected radius. More important, this plot 
shows that this radial segregation of GORES fraction is
modulated by ALOSV: there are significantly more (fewer) GORES among the
higher (lower) ALOSV
galaxies.

\section{Deprojection}

We first deproject the fraction of GORES without taking into account their
ALOSVs.
Writing $g_{\rm GORES}(R) = N_{\rm GORES}(R)/N_{\rm tot}(R)$, we can deduce the
surface densities, $\Sigma(R)=N(R)/(2\pi R)$ of GORES and of all galaxies,
and then perform Abel deprojection yielding space densities
$\nu(r) = -(1/\pi) \int_R^\infty \Sigma'(R)\,dR/\sqrt{R^2-r^2}$ for both GORES
and all galaxies.
The right panel of Fig.~\ref{gGORES} shows that this deprojection leads to a fraction of GORES
that rises linearly with physical radius and then saturates as
\begin{equation}
f_{\rm GORES}(r) = f_0\, {r\over r+a} \ ,
\label{modelfit}
\end{equation}
with best-fit parameters of 
%% where our best fit 
%% to the deprojected fraction of GORES yields 
$f_0=0.52$ and
$a = 1.26\,R_{\rm v}$.

\begin{figure}[ht]
\includegraphics[width=0.49\hsize]{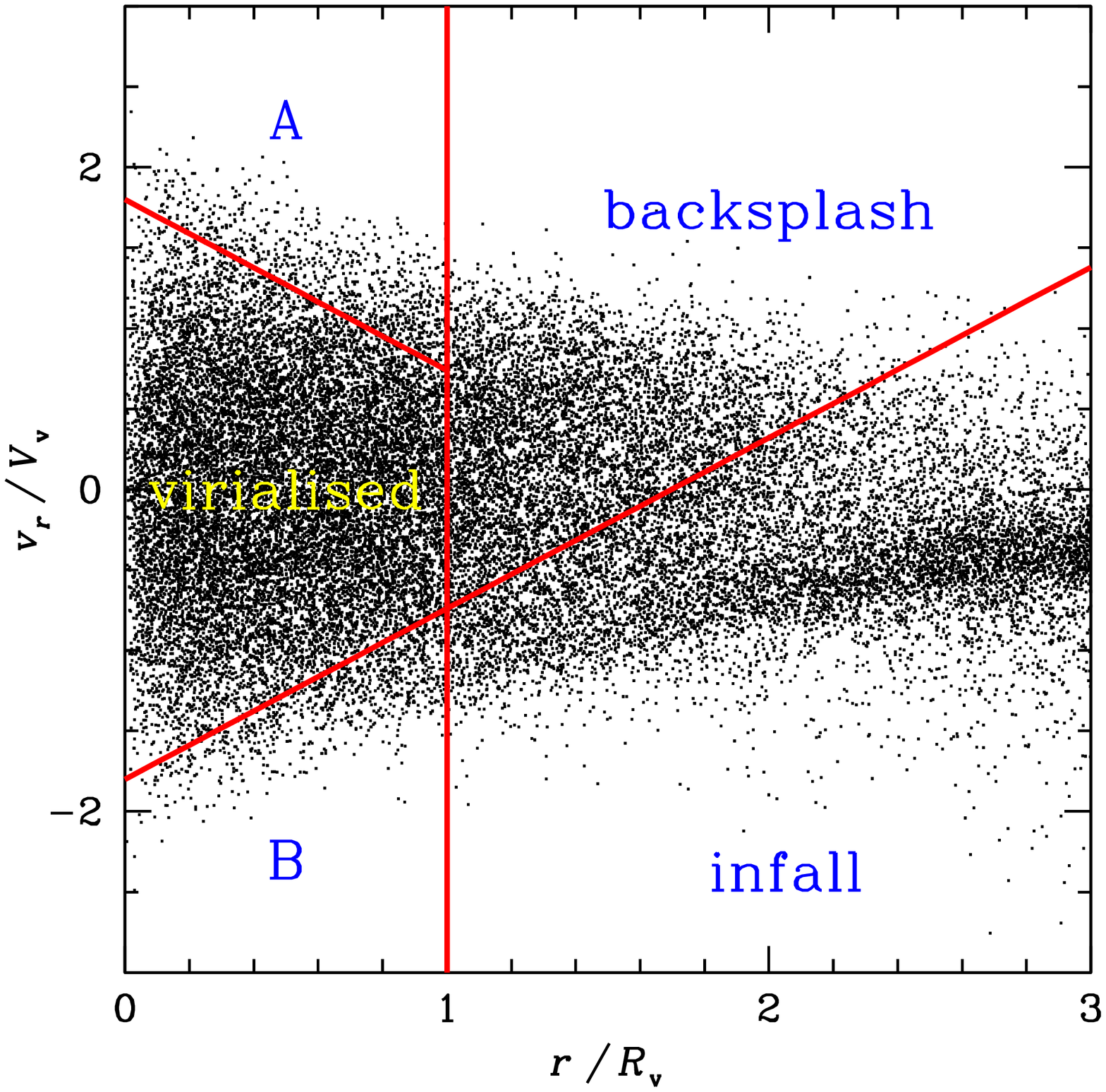}
\includegraphics[width=0.49\hsize]{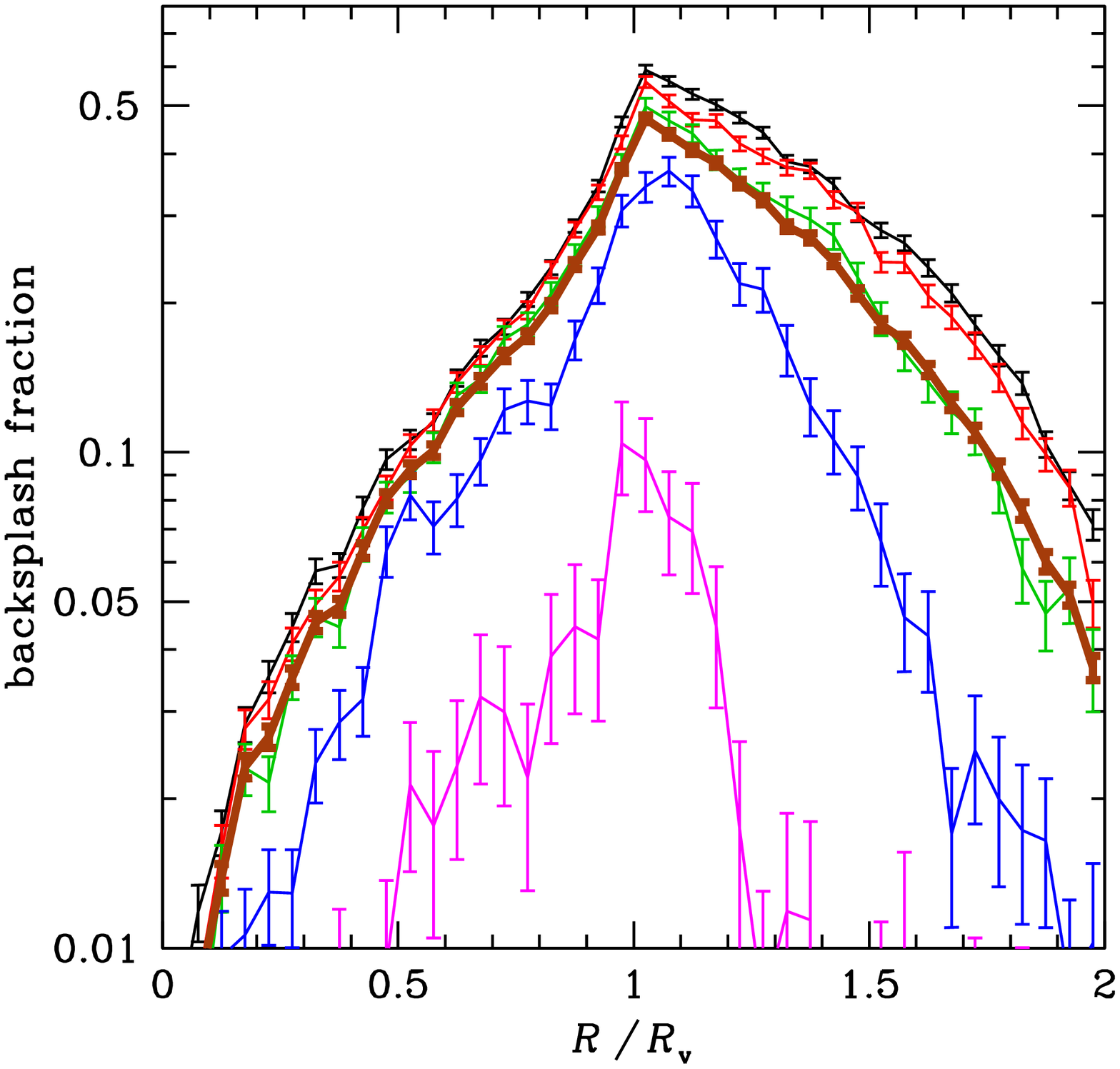}
\caption{\emph{Left}: Radial phase space distribution (in virial units) 
of dark matter particles of a 
stack of 93 mock regular mock clusters from a cosmological hydrodynamical
  simulation \citep{Borgani+04}. 
%% The units of radius and radial velocity
%%   are the virial radius $r_{100}$ and the circular velocity at that radius, respectively.
The critical velocity separating
  infall from backsplash population ({\emph{long diagonal line}}) is 
taken from a single halo of an older, dissipationless, cosmological
simulation \citep{SLM04}. 
%% The {\it{short diagonal line}} represents the
%%   negative critical velocity threshold. 
For
  clarity, only 1 out of 550 particles of the original simulation is plotted.  
The letters indicate uncertain classes.
%% Region A is virialised, backsplash, or infall, while Region B  is virialised in Schemes 0
%%   and 1, backsplash in Schemes 2 and 4 and infall in Scheme 3. 
\emph{Right}: Fraction of backsplash particles in projected phase space, for
our standard scheme, particles in regions A and B are infalling. 
The
\emph{thin lines} indicate ALOSV bins increasing from 0.25, 0.75, ..., 1.75, and
$>2\,\sigma_\upsilon$, going from top to bottom, while the \emph{thick line} shows the average.
}
\label{rvr}
\end{figure}

We now assume that the observed fraction of GORES in projected phase space,
$g_{\rm GORES}(R,|\upsilon_{\,\rm LOS}|)$, is also a fraction of the dynamical 
\emph{class} of
  galaxies: we distinguish virialized, infalling and backsplash galaxies
  according to their $z=0$ position in radial phase space in a cosmological
  simulation as shown in the left panel of Fig.~\ref{rvr}.
The right panel of Fig.~\ref{rvr} shows that in our standard scheme, where
particles in regions A and B are both infalling (this scheme provides the
best fits to the observed fraction of GORES in projected phase space), the
fraction of backsplash particles reaches over 50\% just beyond the virial
radius and for the lowest ALOSVs.

We suppose that the fraction of GORES varies with physical radius $r$ as in
 Eq.~(\ref{modelfit}), varying the normalization and possibly the scale for each class:
\begin{equation}
%f_\alpha \equiv 
f_\alpha(r) = f_\alpha {r/R_{\rm v}\over r/R_{\rm v} + a_\alpha} \ ,
\label{fofr}
\end{equation}
i.e. rising roughly linearly with radius for $r < a_\alpha$, saturating to
an asymptotic value $f_\alpha$ at large radii.
The predicted fractions of GORES is then
\begin{equation}
g_{\rm GORES}(R_i,\upsilon_j) = \sum_\alpha p(\alpha|R_i,\upsilon_j) \sum_k f_\alpha(r_k) \,
q(r_k|R_i,\upsilon_j,\alpha)  \ , 
\label{gvary}
\end{equation}
where
$p(\alpha|R_i,\upsilon_j)$ is the probability that a particle located at projected
radius $R_i$ and ALOSV $\upsilon_j$ is of class $\alpha$, while
$q(k|R_i,\upsilon_j,\alpha)$ is the fraction of the particles of class $\alpha$ in
the cell of
projected 
phase space $(R_i,\upsilon_j)$ that are in the $k$th bin of physical radius
(with $r_k=R_i\,\cosh u_k$, where $u_k$
is linearly spaced from 0 to 
$\cosh^{-1} r_{\rm max}/R_i$, using $r_{\rm max} =
50\,R_{\rm v}$).
We measure $p$ and $q$ from the dark matter particles in the stack of 93
regular mock clusters \citep{MBM10} in a hydrodynamical
cosmological simulation \citep{Borgani+04}.

\begin{figure}[ht]
\includegraphics[width=0.49\hsize]{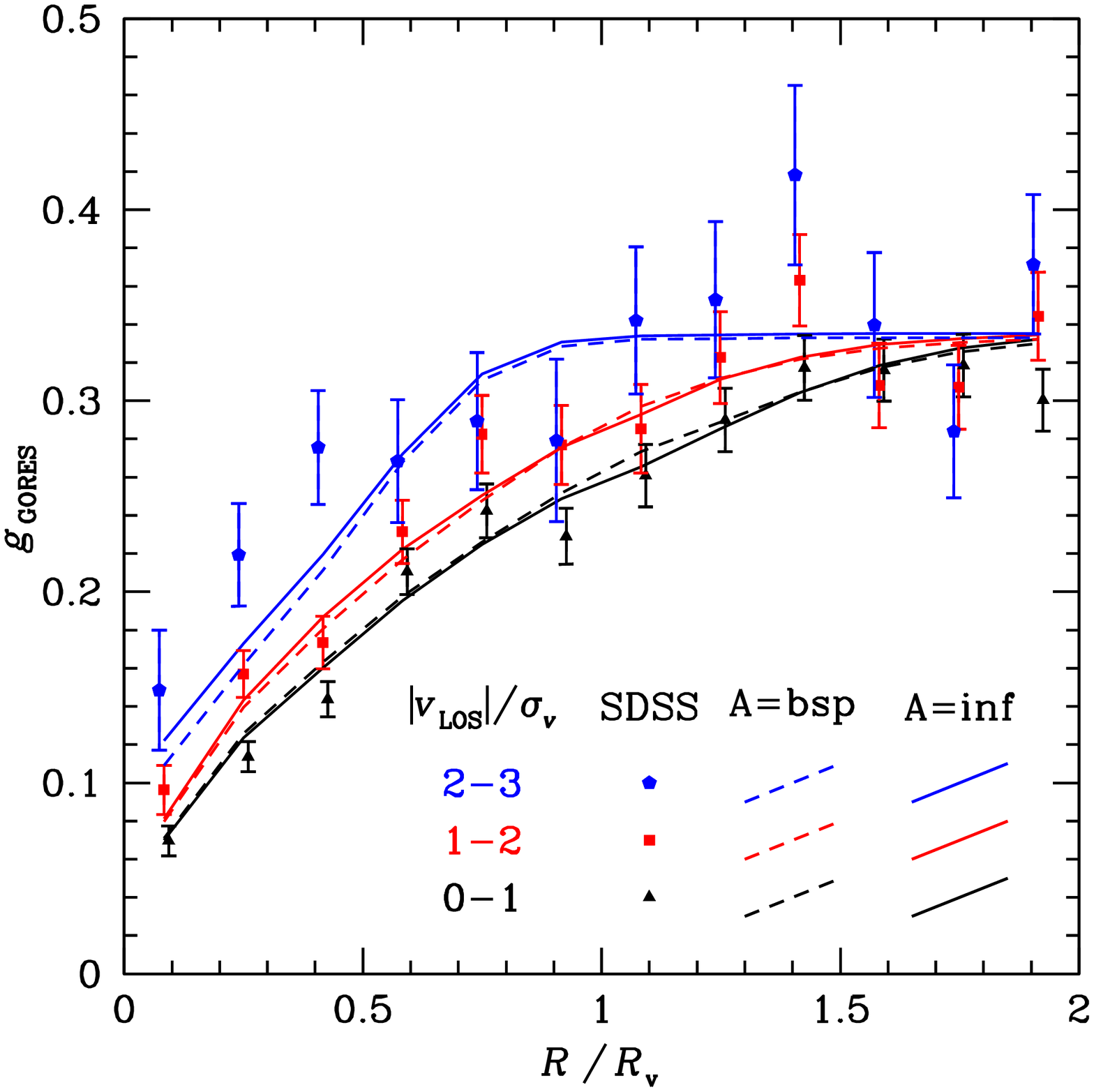}
\includegraphics[width=0.49\hsize]{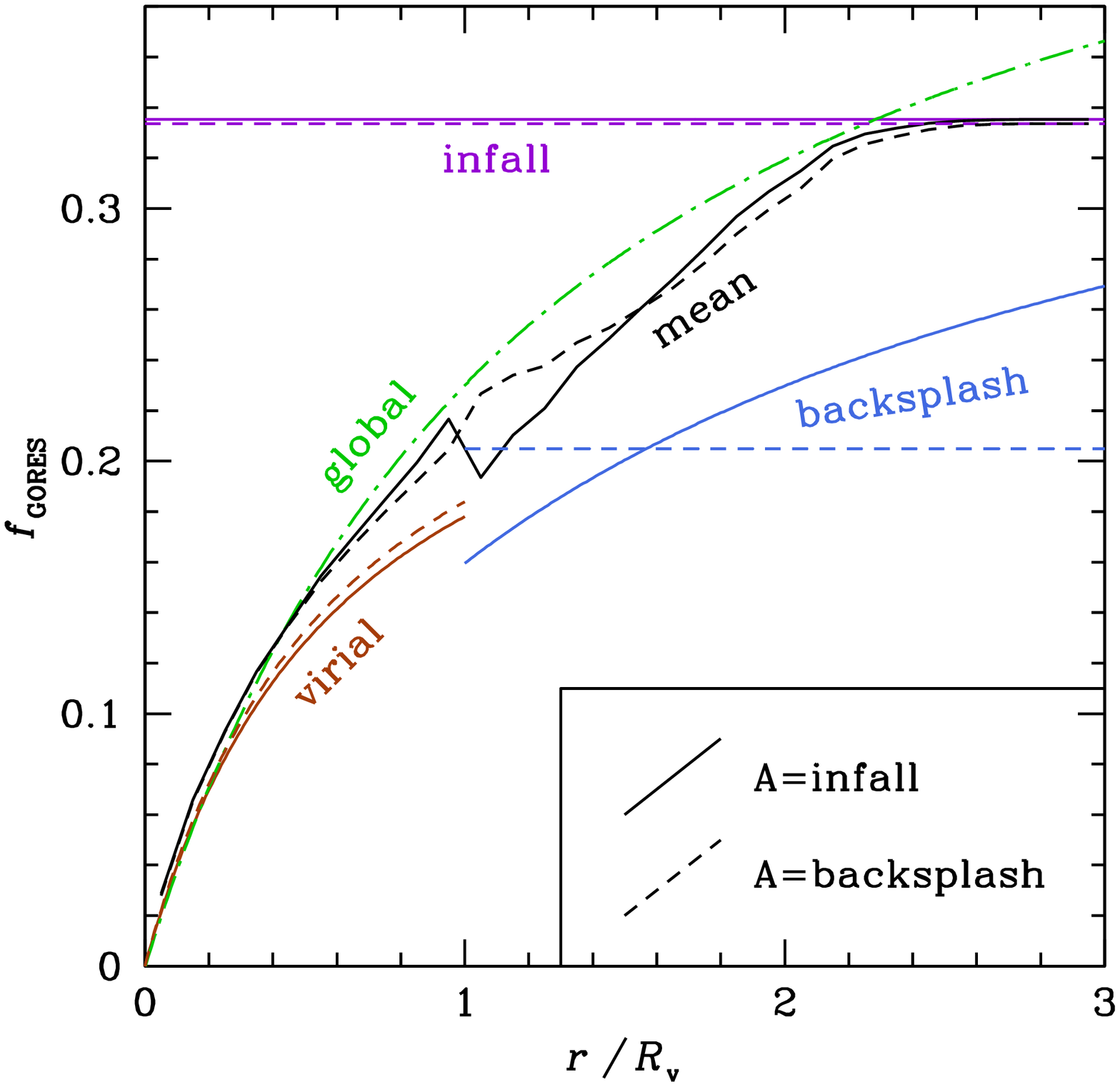}
\caption{\emph{Left}: Observed fraction of GORES with two best-fitting models
  (Eq.~[\ref{fofr}] using Eq.~[\ref{gvary}]) 
  overplotted, both assume that the particles in B are of the infalling
  class. 
\emph{Right}: Deprojected fraction of GORES for the two best-fitting models.}
\label{bestfitplots}
\end{figure}
Fig.~\ref{bestfitplots} shows how the best-fitting models for the two best
fitting schemes (right panel, both with B=infall, but with 
A=infall and A=backsplash, respectively
leading to $\chi^2=1.2$
and 1.4 per degree of freedom with cosmic variance from the simulation included) fit
the observed fraction of GORES in projection (left panel).
Schemes where regions A and B are both virialized or both backsplash, or
where $f_{\rm GORES}$ is constant per class
produce substantially worse fits.

Interestingly, the two best-fitting models yield a GORES fraction that is
independent of radius for the infall population. Moreover, in the best
fitting scheme, the galaxies bouncing out of the cluster with very high
positive line-of-sight velocities are of the infalling class. This suggests
that the quenching of efficient star formation is not instantaneous as the
infalling galaxies pass through the pericenter of their orbit.
Moreover, inspection of the right panel of Fig.~\ref{bestfitplots} indicates
  that at $r$=$R_{\rm v}$, where the three galaxy classes can be compared, 
the backsplash GORES fraction is much closer to that of the virialized class than to the
infall GORES fraction (in fact in our best fitting scheme A=infall, the backsplash
fraction of GORES is lower than the corresponding fraction for the virial
class, but forcing equal fractions of GORES for the two classes at the virial
radius produces an equally good fit).
This suggests that as galaxies cross through clusters, efficient star
formation 
is nearly completely quenched once galaxies reach the virial sphere,
either after the first pericenter on their way out, or at least before they
are mixed with the virialized galaxies.

While GORES account for 18\% of the galaxies within the virial cone, in our
best fitting models, they
also account for 13\% within the virial sphere,  34\% and 19\% of
the infalling and backsplash classes, respectively. GORES also account for as
many as 11\% 
of the virialized galaxies, perhaps the consequence of star formation 
resulting from tides caused by interactions with other galaxies.
% \bibliography{master}

\end{document}